\documentclass[showpacs,showkeys,amsmath,amssymb,superscriptaddress,reprint, preprintnumbers,nofootinbib,notitlepage,aps,prl,twocolumn]{revtex4-1}

\pdfoutput=1 % if your are submitting a pdflatex (i.e. if you have
\usepackage[font=small]{caption}
\usepackage[utf8]{inputenc}
\usepackage[english]{babel}
\usepackage{bm}
\usepackage{graphicx}
\usepackage{dcolumn}
\usepackage{pbox}
\usepackage{epsfig}
\usepackage[dvipsnames]{xcolor}
\usepackage{slashed}
\usepackage{amssymb}
\usepackage{textcomp}
\usepackage{gensymb}
\usepackage{ mathrsfs }
\usepackage{color}
\usepackage[font=small]{caption}
\usepackage[font=small]{subcaption}
\usepackage{url}
\usepackage{comment}
\definecolor{DarkBlue}{rgb}{0.7, 0.4, 1} 
\definecolor{Blue}{rgb}{0, 0.8, 0} 
\definecolor{MyLightBlue}{rgb}{0.5,0.7,1.9}
\definecolor{MyGreen}{rgb}{0.0,0.2, 0.0}
\definecolor{MyBrickRed}{rgb}{0, 0.5, 0.2}
\RequirePackage{hyperref}
\hypersetup{colorlinks, citecolor=Blue,linkcolor=DarkBlue, urlcolor=Green}
\usepackage[normalem]{ulem}

\newcommand{\bea}{\begin{eqnarray}}
\newcommand{\eea}{\end{eqnarray}}

\makeatletter

\renewcommand\@makecaption[2]{%
  \par
  \vskip\abovecaptionskip
  \begingroup
  
   \small\rmfamily
    \begingroup
     \samepage
     \flushing
     \let\footnote\@footnotemark@gobble
     \@make@capt@title{#1}{#2}\par
    \endgroup
  \endgroup
  \vskip\belowcaptionskip
}
\makeatother

\DeclareUnicodeCharacter{2212}{-}
\newcommand{\Mzp}{M_{Z^\prime}}

\newcommand{\mdm}{M_{N_1}}
\newcommand{\gs}{g_\star}
\newcommand{\gss}{g_{\star s}}

\begin{document}
%%%%%%%%%%%%%%%%%%%%%%%%%%%%%%
\title{What KM3-230213A event may tell us about the neutrino mass and dark matter}
%%%%%%%%%%%%%%%%%%%%%
\author{Basabendu Barman}
\email{basabendu.b@srmap.edu.in}
\affiliation{\,\,Department of Physics, School of Engineering and Sciences, SRM University-AP, Amaravati 522240, India}
\author{Arindam Das}
\email{arindamdas@oia.hokudai.ac.jp}
\affiliation{\,\,Institute for the Advancement of Higher Education, Hokkaido University, Sapporo 060-0817, Japan}
\affiliation{Department of Physics, Hokkaido University, Sapporo 060-0810, Japan}
\author{Prantik Sarmah}
\email{prantiksarmah@ihep.ac.cn}
\affiliation{\,\,Institute of High Energy Physics, Chinese Academy of Sciences, Beijing, 100049, People’s Republic of China}
%\preprint{CTPU-PTC-25-06}
%%%%%%%%%%%%%%%%%%%%%%%%%%%
\begin{abstract}   
%%%%%%%%%%%%%%%%%%%%%%%%%%%%%%%%%%
Within the general $U(1)$ scenario, we demonstrate that the ultra high energy neutrinos recently detected by KM3NeT could originate from a decaying right handed neutrino  dark matter (DM), with a mass of 440 PeV. Considering DM production via freeze-in, we delineate the parameter space that satisfies the observed relic abundance, and also lies within the reach of multiple gravitational wave detectors. Our study provides a testable new physics scenario, enabled by multi-messenger astronomy.
%%%%%%%%%%%%%%%%%%%%%%%%%%%%%%%%%%%
\end{abstract}
%%%%%%%%%%%%%%%%%%%%%%%%%%%
\maketitle
%%%%%%%%%%%%%%%%%%%%%%%%%%%%%%%%%%%%%%%%%%%%%%%%%%
\noindent
%%%%%%%%main text%%%%%%%%%%%%%%%%%%%%%%%%%%%%%%%%%
{\textbf{Introduction}.--}
The KM3NeT experiment recently reported the detection of event KM3-230213A~\cite{KM3NeT:2025npi}, involving ultra high energy (UHE) neutrinos in the range $110\,\text{PeV} \leq E_\nu \leq 790\,\text{PeV}$, with a median energy of $220\,\text{PeV}$—the highest-energy neutrino observed on Earth to date. The experiment detected a UHE muon through its deep-sea neutrino telescope, with an energy of $120^{+110}_{-60}\,\text{PeV}$, arriving from an almost horizontal direction $(\rm{RA}: 94.3^{\degree}, \rm{Dec}: -7.8^{\degree})$. This muon is believed to have originated from a more energetic neutrino interacting near the detector. Such energetic neutrinos can be produced in cosmic-ray interactions, specifically via proton-proton and proton-photon collisions at standard astrophysical sources, where  emission of photons associated with neutrinos is guaranteed.  However, no such energetic source is known to exist within the Milky Way or nearby galaxies. The absence of compelling evidence pinpointing the origin of these UHE neutrinos, as observed by KM3NeT~\cite{KM3NeT:2025bxl,KM3NeT:2025aps,KM3NeT:2025vut}, raises the possibility that they have a cosmogenic origin. In particular, they could be produced via the Greisen-Zatsepin-Kuzmin (GZK) effect~\cite{Greisen:1966jv,Zatsepin:1966jv}, in association with UHE gamma rays, as predicted in various models~\cite{Gelmini:2005wu,Aloisio:2015ega,Gelmini:2022evy,Chakraborty:2023hxp}. However, the large uncertainty in the flux of event KM3-230213A—of order $\mathcal{O}(3)$—not only exceeds expectations from current UHE neutrino and gamma-ray flux models but also surpasses the sensitivities of IceCube and the Pierre Auger Observatory (PAO), where no similar event has been observed within a $2.5\sigma$ to $3\sigma$ significance range. This discrepancy challenges the cosmogenic neutrino hypothesis~\cite{Li:2025tqf,KM3NeT:2025ccp}, and calls for other possible explanations.

In~\cite{Barman:2025bir}, we investigated a decaying dark matter (DM) scenario to explain the origin of UHE neutrinos observed by IceCube. The detection of KM3NeT events prompts us to reconsider this framework\footnote{Other possible new physics explanation for KM3NeT events can be found in~\cite{Boccia:2025hpm,Borah:2025igh,Brdar:2025azm,Kohri:2025bsn,Narita:2025udw,Jiang:2025blz,Alves:2025xul,Wang:2025lgn,Yang:2025kfr,Fang:2025nzg,Satunin:2025uui,Dzhatdoev:2025sdi,Neronov:2025jfj,Amelino-Camelia:2025lqn,Crnogorcevic:2025vou,Jho:2025gaf,Klipfel:2025jql,Choi:2025hqt}.}. Our scenario can be naturally embedded within a general anomaly-free $U(1)$ extension of the Standard Model (SM), incorporating three singlet right-handed neutrinos (RHNs)~\cite{Das:2017flq,KA:2023dyz} and a beyond-the-Standard-Model (BSM) scalar singlet. The BSM scalar acquires a nonzero vacuum expectation value (VEV), breaking the $U(1)_{X}$ symmetry and generating Majorana masses for the RHNs, thereby inducing the seesaw mechanism. It gives rise to tiny neutrino masses and flavor mixing~\cite{Minkowski:1977sc,Yanagida:1979as,Gell-Mann:1979vob,Mohapatra:1979ia,Schechter:1980gr,Sawada:1979dis,Mohapatra:1980yp}. We consider a long-lived decaying DM candidate, that can be identified as the lightest RHN with a mass in the PeV range, depending on the neutrino mass hierarchy—either normal (NH) or inverted (IH). By fitting neutrino oscillation data~\cite{ParticleDataGroup:2024cfk}, we estimate the DM lifetime for both NH and IH cases and use this to reproduce the expected neutrino and photon fluxes at KM3NeT~\cite{KM3NeT:2025bxl,KM3NeT:2025aps,KM3NeT:2025vut}. This PeV-scale decaying DM, which explains the origin of UHE neutrino event with $\langle E_\nu\rangle \simeq 220$ PeV, could be produced via freeze-in mechanism~\cite{Hall:2009bx}. Interestingly, in~\cite{Boyle:2018tzc}, the author's argued that if the Universe
is in its preferred CPT-symmetric vacuum, then such a PeV-scale DM naturally emerges from Big Bang like Hawking radiation from a black hole. The breaking of $U(1)_{X}$ symmetry also leads to the formation of one-dimensional topological defects in the form of cosmic strings (CS)~\cite{Nielsen:1973cs,Kibble:1976sj}, characterized by a string tension $G\mu \sim \mathbb{B}\,G v^2_\Phi$, where $v_\Phi$ is the VEV of the $U(1)_{X}$ symmetry-breaking singlet scalar, and $\mathbb{B} \sim 0.1$~\cite{Babul:1987me,Vilenkin:2000jqa,Dror:2019syi}. Thus, this scenario provides a compelling framework in which the KM3NeT event not only provides a hint towards decaying DM, but sheds light on the neutrino mass hierarchy as well. Moreover, the gravitational waves (GWs) from cosmic strings predicted in this scenario could be probed by several proposed GW detectors.  Finally, we constrain very heavy $Z^\prime$, that acquires mass via spontaneous breaking of $U(1)_X$, from KM3-230213A events, DM abundance and GW from CS. Such heavy gauge bosons are beyond the scope of existing direct search experiments.
\\

%%%%%%%%%%%%%%%%%%%
\noindent
{\textbf{The model framework}--}
%%%%%%%%%%%%%%%%%%%%%%%%%%%%%%%%%%%%%%%%%%%%%%%%
Under the $\text{SM}\otimes U(1)_{X}$ gauge symmetry, the SM quarks transform as $q_L^i=\{3,2,\frac{1}{6}, \frac{1}{6}x_H+\frac{1}{3} x_\Phi\}$, $u_R^i=\{3,1,\frac{2}{3}, \frac{2}{3}x_H+\frac{1}{3} x_\Phi\}$, $d_R^i=\{3,1,-\frac{1}{3}, -\frac{1}{3}x_H+\frac{1}{3} x_\Phi\}$, respectively. The SM leptons transform as $\ell_L^i=\{1,2,-\frac{1}{2}, -\frac{1}{2} x_H- x_\Phi\}$, $e_R^i=\{1,1,-1, -x_H-x_\Phi\}$, respectively, while the SM Higgs transforms as $H=\{1,2,\frac{1}{2}, \frac{x_H}{2}\}$ with $\{x_H,\, x_\Phi\}\in\Re$. Three SM-singlet RHNs, required to cancel gauge and mixed gauge-gravity anomalies, transform as $N_R^{i}=\{1,1,0,-x_\Phi\}$ with $i=1, 2, 3$ and one SM-singlet $U(1)_{X}$ scalar transforms as $\Phi=\{1,1,0, 2 x_\Phi\}$. The Yukawa interactions relevant for the neutrino mass can be written as
\bea
{\cal L} &\supset& - Y_{\nu_{\alpha \beta}} \overline{\ell_L^\alpha} \tilde{H}\,N_R^\beta- \frac{1}{2}Y_{N_\beta} \Phi \overline{(N_R^\beta)^c} N_R^\beta + {\rm H.c.,} 
\label{eq:LYk}   
\eea
where we consider a basis where $Y_{N_{\alpha}}$ is a diagonal matrix, and $\tilde{H} = i \tau^2 H^*$ with $\tau^2$ being the second Pauli matrix. The scalar potential of this scenario is given by 
\bea
V=\sum_{\mathcal{I}= H, \Phi} \Big[m_{\mathcal{I}}^2 (\mathcal{I}^{\dagger} \mathcal{I})+ \lambda_{\mathcal{I}} (\mathcal{I}^\dagger \mathcal{I})^2 \Big] +
\lambda_{\rm mix} (H^\dagger H) (\Phi^{\dagger} \Phi)\,.
\label{pot}
\eea 
After the breaking of $U(1)_X$ and electroweak gauge symmetries, the scalar fields $H$ and $\Phi$ develop their VEVs as 
\begin{align}\label{eq:VEV}
  \langle H\rangle \ = \ \frac{1}{\sqrt{2}}\begin{pmatrix} v+h\\0 
  \end{pmatrix}~, \quad {\rm and}\quad 
  \langle\Phi\rangle \ =\  \frac{v_\Phi^{}+\phi}{\sqrt{2}}~,
\end{align}
where electroweak scale is  $v=246$ GeV at the potential minimum. In the limit $v_\Phi\gg v$, the mass of the $U(1)_X$ gauge boson can be written as $M_{Z^\prime}^{}= 2 g_X x_\Phi v_\Phi$. Taking $x_\Phi=1$ without the loss of generality we find that for $x_H=-2$, $\ell_L$ and $q_L$ do not have interactions with $Z^\prime$, for $x_H=0$ left and right handed SM fermions interact equally with $Z^\prime$ which provides the B$-$L scenario and finally for $x_H=-1$, $e_R$ does not interact with $Z^\prime$. The breaking of $U(1)_X$ and electroweak symmetries induce the Majorana and Dirac mass terms for RHNs and  light left-handed neutrinos from Eq.~\eqref{eq:LYk} as
\begin{equation}
M_{N_\beta}^{} \ = \ \frac{Y_{N_\beta}}{\sqrt{2}} v_\Phi^{}, \,\,\,
m_{{D}_{\alpha \beta}} \  =  \ \frac{Y_{{\nu}_{\alpha \beta}}}{\sqrt{2}} v.
\label{eq:mDI}
\end{equation}
Hence light active neutrino masses follow the seesaw formula 
\bea
m_\nu\simeq -m_D^{} M_{N}^{-1} m_D^T\,,
\label{nm}
\eea
explaining the origin of tiny neutrino masses and flavor mixing. Diagonalizing Eq.~(\ref{nm}) one obtains
\bea
\mathcal{U}^T m_\nu \mathcal{U} ={\rm diag}(m_1, m_2, m_3)\,.
\eea
%%%%%%%%%%%%%%%%%%%%%%%%%%
where $\mathcal{U}$ is the PMNS matrix and it depends on neutrino oscillation data \cite{ParticleDataGroup:2024cfk}. We consider the light neutrino mass eigenvalues  follow $m_1=m_{\rm lightest} < m_2 < m_3$ in NH and  $m_3=m_{\rm lightest} < m_1< m_2$ in IH cases where $m_{\rm lightest}$ is the lightest light neutrino mass being a very small free parameter (see Sec.~\ref{sec:nu-mass} for details). From the seesaw formula we write 
\bea
V_{\alpha \beta}^{\rm{NH(IH)}} \ = \mathcal{U}_{}^{\ast} \sqrt{D^{\rm{NH(IH)}}} \sqrt{M_N^{-1}},
\label{gp1}
\eea
when $M_N$ is in diagonal basis for NH (IH) case. From the definition of neutrino mixing we obtain
\bea
V_{\alpha \beta}^{\rm{NH(IH)}}=m_{D_{\alpha \beta}}^{\rm{NH(IH)}}/M_{N_\beta}=
\frac{Y_{\nu_{\alpha \beta}}^{\rm{NH(IH)}} v}{\sqrt{2} M_{N_\beta}}.
\label{mix}
\eea

Out of three RHN species, we identify the lightest RHN species ($N_{1(3)}$ in NH(IH)) as a long-lived decaying DM candidate ensured by the Yukawa coupling $Y_{\nu_{\alpha \beta}}^{\rm NH(IH)}$. Partial decay widths of very heavy RHNs \cite{Das:2016hof} neglecting the masses of SM bosons are 
\begin{align}
&\Gamma (N_\beta \rightarrow \ell_\alpha W)\simeq \frac{\lvert Y_{\nu_{\alpha \beta}}^{\rm NH(IH)}\rvert^2 M_{N_\beta}}{32 \pi},\nonumber \\&
\Gamma (N_\beta \rightarrow \nu_\alpha Z)= \Gamma (N_\beta \rightarrow \nu_\alpha h) \simeq \frac{\lvert Y_{\nu_{\alpha \beta}}^{\rm NH(IH)}\rvert^2 M_{N_\beta}}{64 \pi}\,,
\label{decay}
\end{align}
while its radiative decay rate is given by~\cite{Pal:1981rm,Shrock:1982sc} 
\begin{align}
&\Gamma_{N_\beta\rightarrow \nu_\alpha\,\gamma} \simeq M_{N_\beta}\,\frac{\alpha_{\rm EM}\,\lvert Y_{\nu_{\alpha \beta}}^{\rm NH(IH)}\rvert^2}{256\,\pi^6}\,\left(\frac{m_\mu}{v}\right)^2\,,
\end{align}
where $\alpha_{\rm EM}=1/137$ is  the fine structure and Fermi constants, respectively. However, the radiative decay rate has a sub-dominant contribution to the total decay rate. 

In the context of KM3-230213A, $N_{1(3)}$ could be a potential DM candidate for the NH (IH) case. The DM lifetime, including all possible two-body decays, reads
\begin{equation}
\tau_{\rm{DM}}^{\rm NH(IH)} \approx (7.5\times 10^{30}\,{\rm s})\,\left(  \frac{440\, {\rm PeV}}{M_{1(3)}}\right)\,\left (\sum_\alpha \bigg \lvert \frac{Y_{\nu_{\alpha 1(3)}}^{\rm NH(IH)}}{10^{-31}} \bigg \rvert^2 \right)^{-1}\,.
\label{lifetimedm}
\end{equation}
%%%%%%%%%%%%%%%%%%%%%%%
\begin{figure*}
    \centering    \includegraphics[width=0.5\linewidth]{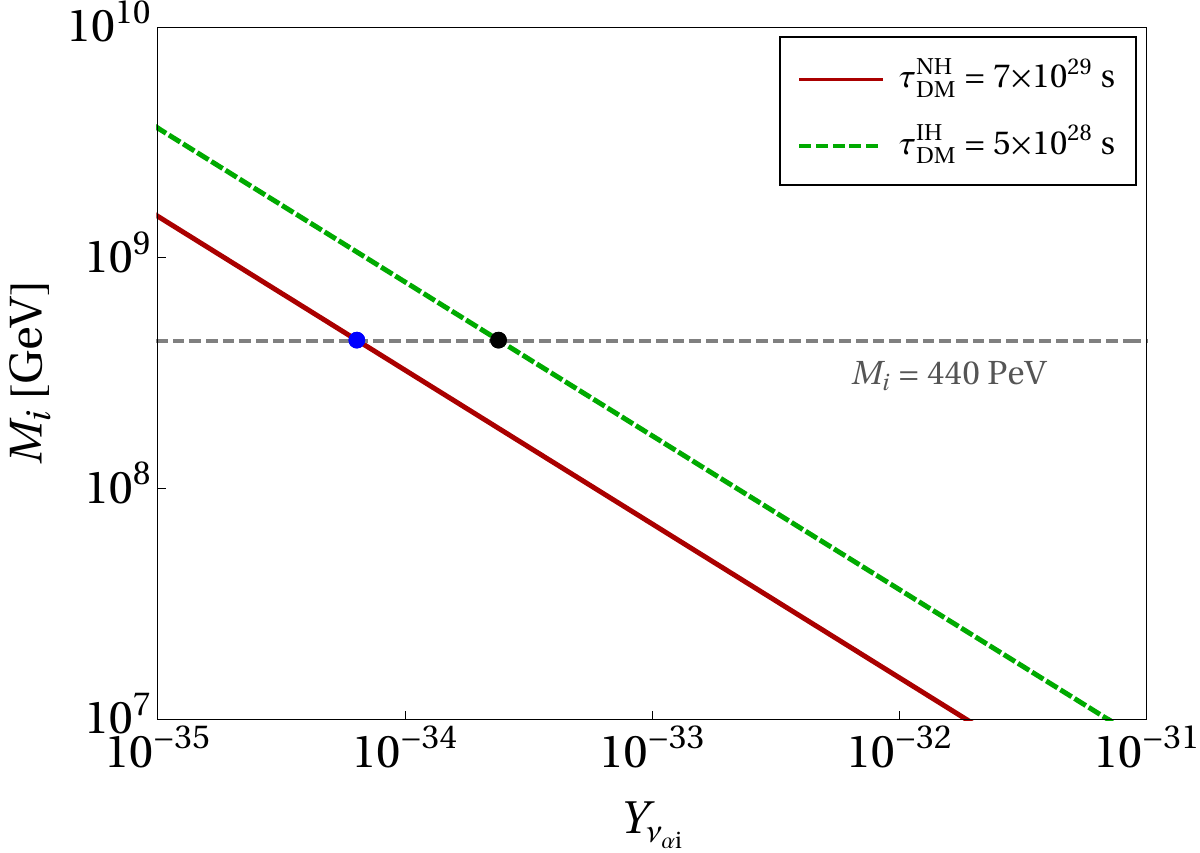} 
    \caption{Contours of DM lifetime, corresponding to normal (solid, red) and inverted (dashed, green) hierarchy. The blue and black points correspond to the Yukawas mentioned in Tab.~\ref{tab:bp}. The horizontal dashed line denotes the DM mass required to explain the KM3NeT event.}
    \label{fig:dmlife}
\end{figure*}
%%%%%%%%%%%%%%%%%%%%%%%%
Fitting neutrino oscillation data \cite{ParticleDataGroup:2024cfk} in Eqs.~(\ref{gp1}) and (\ref{mix}) we find Yukawa couplings $Y_{21}^{\rm NH} (Y_{23}^{\rm IH})\simeq 3.27\times 10^{-31}\, (1.22\times 10^{-30})$ between $N_{1(3)}$ and $\nu_\mu$. We defined the lightest light neutrino mass as $m_{\rm lightest}=r\,\sqrt{\Delta m_{12}^2}=4.39\times 10^{-65}\,\left(2.11\times 10^{-64}\right)$ GeV for NH (IH) where $r$ is a tiny free parameter of benchmarks values $5.03\times 10^{-54}\,(2.42\times10^{-53})$ for NH (IH) case. Using $Y_{21}^{\rm NH}\, (Y_{23}^{\rm IH})$ in Eq.~\eqref{lifetimedm} we find $\tau_{\rm DM}^{\rm NH (IH)}=7 \times 10^{29}\,(5\times 10^{28})$s. The details of the benchmark pints chosen above are tabulated in Tab.~\ref{tab:bp}. 
%%%%%%%%%%%%%%%%
\begin{table*}[htb!]
\centering
\setlength{\tabcolsep}{8pt} % Increases column separation (default is 6pt)
\renewcommand{\arraystretch}{1.5} % Increases row height (default is 1)
\begin{tabular}{|c|c|c|c|} 
\hline
$Y_{21}^{\rm NH}$ & $Y_{23}^{\rm IH}$ & $\tau_{\rm DM}^{\rm NH}$ [s] & $\tau_{\rm DM}^{\rm IH}$ [s]\\ 
\hline\hline
$3.27\times 10^{-31}$ & $1.22\times 10^{-30}$ & $7 \times 10^{29}$ & $5\times 10^{28}$
\\
\hline
\end{tabular}
\caption{Benchmark values of active-sterile Yukawa coupling along with corresponding DM decay lifetime, for normal and inverted hierarchies. The DM mass is fixed to 440 PeV.} 
\label{tab:bp}
\end{table*}
%%%%%%%%%%%%%%%%

In Fig.~\ref{fig:dmlife} we show DM lifetime as a function of the Yukawa $Y_\nu$, for the two benchmark points in Tab.~\ref{tab:bp}. Since $\Gamma_{\rm DM}\propto M_N\,|Y_\nu|^2$, hence a larger coupling requires smaller $M_N$ to obtain a constant lifetime. It is crucial to mention that the long-lived RHN, which is a potential DM candidate, does not participate in the neutrino mass generation mechanism. 
\\

%%%%%%%%%%%%%%%%%%%%%%%%%%%%%%%%%%
\noindent
{\textbf{Dark matter induced explanation of the excess}--}
%%%%%%%%%%%%%%%%%%%%%%%%%%%%%%%%%%%%%%%%%%%%%%%%%%%%%%%%%%
The decay of $N_{1(3)}$ can yield high energy gamma-rays and neutrinos in the final state. To compute the flux of these secondaries from $N_{1(3)}$ in the Milky Way (MW), we choose the  the extensively explored  Navarro-Frenk-White (NFW)~\cite{Navarro:1996gj,Navarro:2003ew} density profile of the Milky way DM halo among many modes of DM profile ~\cite{Navarro:1996gj,Navarro:2003ew,1989A&A...223...89E,Graham:2005xx,Burkert1995}. As a function of the 
Galactocentric radius $(R_{\rm GC})$ the NFW profile becomes,
\begin{equation}
  \rho_{\rm DM} =  \rho_{\rm NFW}(R_{\rm GC}) = \frac{\rho_{\rm C}}{(R_{\rm GC}/R_{\rm C})(1+R_{\rm GC}/R_{\rm C})^2} \ ,
\end{equation}
where, $R_{\rm C} = 11$~kpc and $\rho_{\rm C}$ are the characteristic scale and density, respectively where $\rho_{\rm C}$  is obtained by normalizing the DM profile to the DM density at solar neighbourhood, $\rho_{\odot}= 0.43~\rm GeV cm^{-3}$. The flux of secondary gamma-rays and neutrinos can be obtained by estimating the amount of DM  in the angular region, $\Delta \Omega$ of observation and given by,
\begin{equation}
    \mathcal{D}  = \frac{1}{\Delta \Omega} \int_{\Delta \Omega} \mathrm{d} \Omega \int_{0}^{s_{\rm max}} \mathrm{d}s\,\rho_{\rm DM} (s, b, l). 
    \label{eq:Flux-formula}
\end{equation}
Here, $s$ is the line of sight distance which is connected to the  Galactic longitude ($l$) and latitude ($b$) by the relation $r=\sqrt{s^2 + R_{\odot}^2 + 2 s R_{\odot} \cos{b}\cos{l}}$, where $R_{\odot}=8.3~\rm kpc$ is the distance to the Milky Way center from the Sun. The flux of the secondaries is given by,
\begin{equation}
    \frac{\mathbf{d}^2 \phi_{i, \rm G} (E_{i})}{\mathrm{d} E_{i} \mathrm{d} \Omega} = \frac{\mathcal{D}}{4 \pi M_{\rm DM} \tau_{\rm DM} }  \frac{\mathrm{d} N_{i}(E_i)}{\mathrm{d}E_{i}}\,,
        \label{eq:Flux-formula1}
\end{equation}
where, $i$ represents gamma-rays or neutrinos of any specific flavor. In addition to the Galactic DM, extra-galactic DM can also contribute to the high-energy gamma-ray and neutrino fluxes, which can be obtained by the following formula,
\begin{align} \label{eq:eg}
    \frac{d \phi_{i, \rm EG} (E_i)}{dE_i} = \frac{c\, \rho_{\rm DM}}{4\pi\, M_{\rm DM }  \tau_{\rm DM}} \int dz \left|\frac{dt}{dz}\right| \frac{dN (E_i')}{dE_i'} e^{-\tau_{\rm OD} (E_i^{\prime},z)}\,,
\end{align}
where $E_i' = E_i(1+z)$ corresponds to the energy of the $i^{\rm th}$ particle at redshift $z$. The DM density is given by $\rho_{\rm DM}= \Omega_{\rm DM} \rho_{c}$, where $\rho_{c}=4.7\times 10^{-6}$ GeV cm$^{-3}$ is the critical DM density in a flat Friedmann–Lema${\rm \hat{i}}$tre–Robertson–Walker (FLRW) Universe and $\Omega_{\rm DM}=0.27$. The cosmological line element, $ \left|\frac{dt}{dz}\right|$ can be expressed as 
\begin{eqnarray}\label{eq:eg1}
     \left|\frac{dt}{dz}\right| = \frac{1}{\mathcal{H}_{0}(1+z)\sqrt{(1+z)^3 \Omega_{\rm m} + \Omega_{\Lambda}}},
\end{eqnarray}
where $\Omega_{m}=0.315$, $\Omega_{\Lambda}=0.685$ and $\mathcal{H}_{0} = 67.3$ km s$^{-1}$ Mpc$^{-1}$ is the current expansion rate of the Universe. The  factor $e^{-\tau_{\rm OD} (E_{i}^{\prime},z)}$ takes into account of the attenuation of the 
gamma-ray flux due to pair production losses in the extra-galactic background light (EBL) and CMB  during propagation, where $\tau_{\rm OD} (E_{i}^{\prime},z)$ is the total optical depth of EBL and CMB. The optical depth (OD) of EBL is taken from ~\cite{Stecker:2016fsg}. The term $\mathrm{d} N_{i}/ \mathrm{d} E_i$ in Eq.~\eqref{eq:Flux-formula1} and Eq.~\eqref{eq:eg} represents differential spectra of secondary gamma-rays and neutrinos as a function of energy, $E_i$ and obtained from %the publicly available code 
\texttt{HDMSpectra}~\cite{Bauer:2020jay}. 

\begin{figure*}
    \centering    \includegraphics[width=0.7\linewidth]{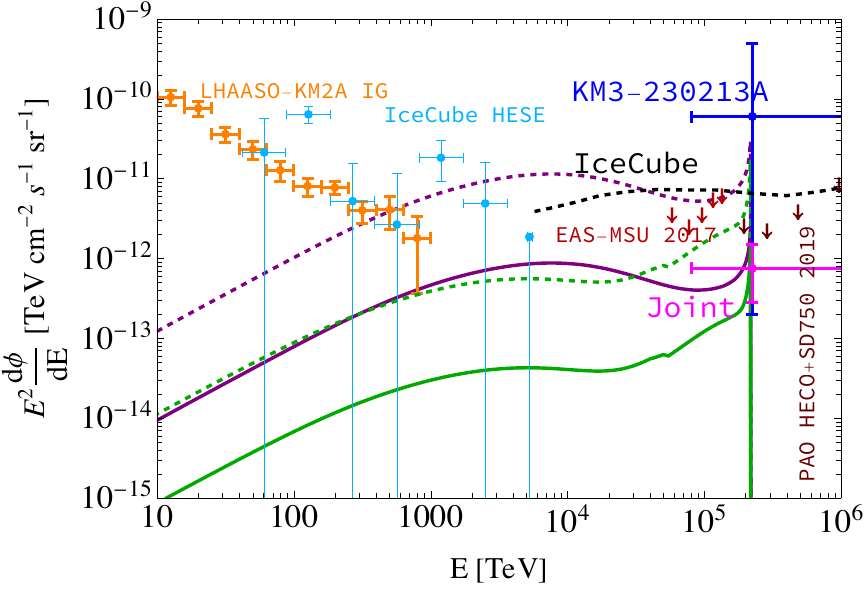} 
    \caption{Gamma-ray (dashed curve) and $\nu_{\mu}$ (solid curve) fluxes for $\tau_{\rm DM}= 7 \times 10^{29}\,\text{s}$ (in green, NH) and $5\times 10^{28}\,\text{s}$ (in purple, IH), respectively, considering $M_{\rm DM}=440$ PeV. We compare our results with the fluxes from KM3NeT~\cite{KM3NeT:2025npi} event with $3 \sigma$ error bars (blue), best fit flux from joint analysis of KM3NeT + IceCube + PAO (magenta),  IceCube HESE~\cite{IceCube:2020wum} data (light blue), $90\%$ CL sensitivity of IceCube~\cite{IceCube:2018fhm} for cosmogenic neutrinos (black dotted), measurement of diffuse gamma-rays from the inner Galactic (IG) plane by LHAASO~\cite{LHAASO:2023gne} (orange), the upper limits on UHE gamma-ray flux by EAS-MSU~\cite{Fomin:2017ypo} (maroon downward arrow) and PAO HECO +SD750~\cite{Castellina:2019huz} (brown downward arrows), respectively.}
    \label{fig:flux}
\end{figure*}

The total flux ($\phi_{i}$), obtained by summing over both the Galactic and extra-galactic components, depends on two free parameters: (i) the (decaying) DM mass $M_{\rm DM}$ being set at 440 PeV to kinematically allow the secondaries acquire energy about 220 PeV from 2-body decay and (ii) the DM lifetime $\tau_{\rm DM}$ where we consider gamma-ray observations by different telescopes across a broad energy band, $(10^{1}-10^{6})$ TeV. In the lower energies, $E_{\gamma} < 10^{3}$~TeV, we adopt the diffuse flux measurements from the inner Galactic plane ($15^{\circ}<l<125^\circ$ and $-5^{\circ}<b<5^{\circ}$) by  the Kilometer Square Array at the Large High Altitude Air Shower Observatory (LHAASO-KM2A)~\cite{LHAASO:2023gne}. For higher energies, we constrain the gamma-ray flux from DM decay using the upper limits on UHE photon flux from Moscow State University Extensive Air Shower (EAS-MSU) array~\cite{Fomin:2017ypo} and Pierre Auger Observatory  (PAO HECO +SD750)~\cite{Castellina:2019huz}. The gamma-ray flux calculated for the inner Galactic plane from the $N_{1(3)}\to \nu_\mu \gamma$ decay assuming  NH (IH) is represented by the solid (dashed) purple curve in Fig.~\ref{fig:flux}. 
For comparison, we include gamma-ray data from LHAASO-KM2A (orange), EAS-MSU (maroon downward arrows), and PAO HECO + SD750 (brown downward arrows). To remain consistent with these observations, $\tau_{\rm DM}^{\rm NH (IH)}$ is set to $7 \times 10^{29}(5 \times 10^{28})~\rm s$. The $\nu_\mu$-flux from $N_{1(3)}$ decay, computed for the angular region constrained by the uncertainty in the arrival direction of KM3-230213A, is shown as the solid (dashed) green curve. We account for neutrino oscillations during propagation, leading to a final neutrino flavor ratio at Earth of $3:2:1$ for NH and $1:2:3$ for IH~\cite{Athar:2000yw}. Consequently, the muon neutrino flux after oscillations remains similar in both cases. While the predicted neutrino flux is over an order of magnitude lower than the central flux value of the KM3-230213A event, it still lies within its $3\sigma$ uncertainty. Thus, a DM origin for the KM3-230213A event cannot be completely ruled out. For consistency with existing observations, we also display IceCube's High Energy Starting Events (HESE)~\cite{IceCube:2020wum} as cyan data points. While the resulting neutrino flux is close to the central value of KM3-230213A, the corresponding gamma-ray flux is in tension with the observed gamma-ray data. It is important to note that the choice of $\tau_{\rm DM}$ is not unique, by adjusting the Yukawa parametrization, one can fine-tune the gamma-ray flux to match observations.  
%%%%%%%%%%%%%%%%%%%%%%
%%%%%%%%%%%%%%%%%%%%%%%

\begin{figure*}
    \centering    \includegraphics[width=0.47\linewidth]{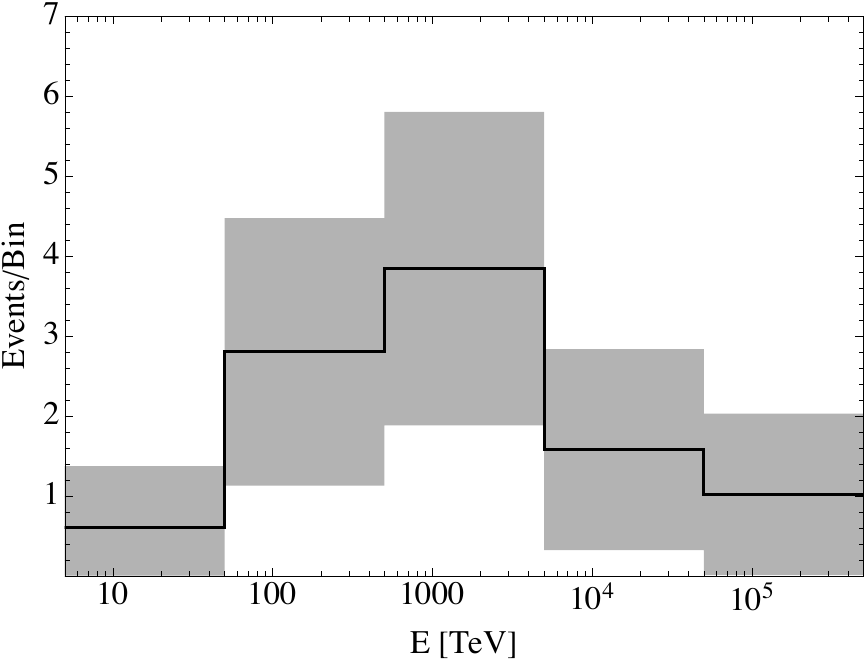}    \includegraphics[width=0.51\linewidth]{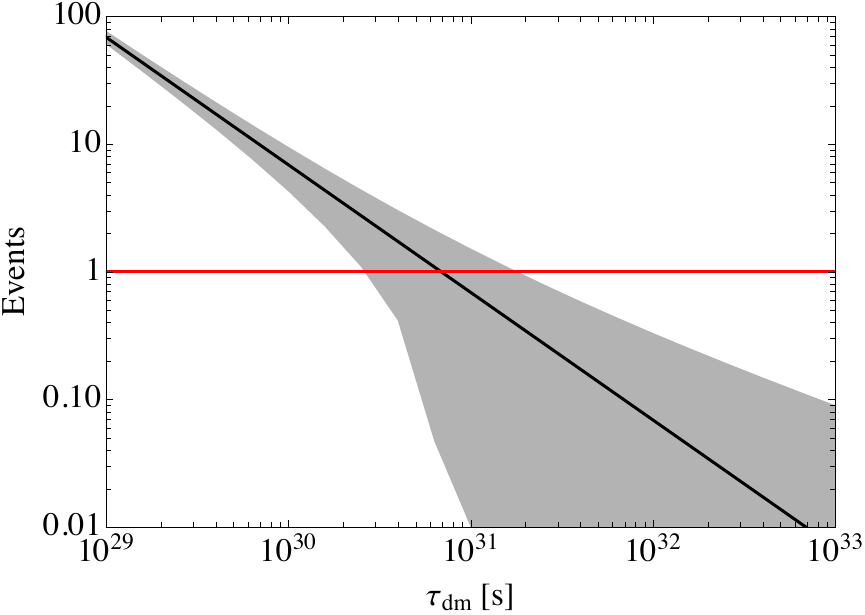}
    \caption{Left: Expected number of muon neutrino events in KM3NeT for a observation time of 10 years. Right: The gray band shows the $1 \sigma$ uncertainty in the events in Poisson statistics. The red straight line parallel to the horizontal axis marks 1 event counted by KM3NeT.}
    \label{fig:events}
\end{figure*}

We also estimate the expected number of muon neutrino events at KM3NeT for our predicted neutrino flux from DM decay. The number of events $N_{\rm event}$ in the energy interval $\Delta E_{\nu_{\mu}}$ is given by,
\begin{equation}
    N_{\rm event} =\mathcal{T}_{\rm obs}\int_{E_{\nu_{\mu}}}^{E_{\nu_{\mu}}+\Delta E_{\nu_{\mu}}}  \mathrm{d}E_{\nu_{\mu}}\, \frac{\mathrm{d \phi_{\nu_{\mu}}}}{\mathrm{d} E_{\nu_{\mu}}}\,\mathcal{A}_{\rm eff}(E_{\nu_{\mu}})\,,
\end{equation}
where $\mathcal{T}_{\rm obs}$ is the observation time and $\mathcal{A}_{\rm eff}(E_{\nu_{\mu}})$ is the effective area of KM3NeT detector for muon neutrinos~\cite{KM3NeTWebsite}. For the computation of the events, we consider the diffuse flux integrated over the entire Galactic DM halo. In the left panel of Fig.~\ref{fig:events}, we show the expected number of events (black line) as a function energy for a observation time of 10 years. The gray band shows the $1\sigma$ spread in the events considering Poisson statistics, i.e., $\sigma =\sqrt{N_{\rm event}}$. In addition, we show the variation of the total events integrated in the energy range $5$~TeV to $5 \times 10^{5}$~TeV with respect to $\tau_{\rm dm}$ in the right panel. The gray band corresponds to the $1\sigma$ uncertainty, and the red horizontal line corresponds to $1$ event. As it is evident, KM3NeT will be able to probe $\tau_{\rm dm}$ up to around $10^{31}$~s after 10 years of observation. 

It is important to mention that future detection of additional diffuse neutrinos by KM3NeT, IceCube, Auger, and upcoming experiments---such as IceCube-Gen2, GRAND, HUNT, and TRIDENT---could significantly reduce this uncertainty and help resolve the current tension. Indeed, a recent joint analysis~\cite{KM3NeT:2025ccp} by the KM3NeT, IceCube, and PAO collaborations has reported a revised flux  estimate (magenta cross in Fig.~\ref{fig:flux}) that is approximately two orders of magnitude lower than the initial one. While this mitigates the discrepancy, a residual tension at the $(2.5$--$3)\sigma$ level persists. Notably, this lower flux level is compatible with a DM interpretation, as illustrated by the solid green curve in Fig.~\ref{fig:flux}. Further detections will enable tighter constraints on the neutrino flux and, by extension, on models of DM origin. The energy spectrum of the observed neutrinos is particularly critical for disentangling astrophysical and non-astrophysical sources. Astrophysical sources are typically expected to produce a power-law spectrum, often correlated with UHE cosmic rays and UHE gamma rays. 
In contrast, DM-induced spectra can exhibit distinct features, such as the spectral spike shown in Fig.~\ref{fig:flux}, which is generally not anticipated from conventional astrophysical sources. Nevertheless, future detection of more UHE neutrinos across broad energy range will help resolve these UHE sources.\\

%%%%%%%%%%%%%%%%%%
%%%%%%%%%%%%%%%%%%%%%%%

%%%%%%%%%%%%%%%%%%%%
\noindent
{\textbf{Dark matter genesis}--}
%%%%%%%%%%%%%%%%%%%
DM $N_{1(3)}$ at PeV-scale can be produced via freeze-in through the following mechanisms:  
(i) the on-shell decay of $Z^\prime$, provided that $M_{Z^{\prime}} > 2\,M_{1(3)}$;  (ii) the on-shell decay of $\phi$, if $m_\phi > 2\,M_{1(3)}$; and  (iii) $2 \to 2$ scattering of thermal bath particles mediated by $Z^\prime$ (see Appendix.~\ref{sec:decay} and \ref{sec:cs} for detailed expressions). The coupling strength $g_X$ must be sufficiently small to ensure non-thermal DM production via freeze-in. Consequently, $Z^\prime$ never attains thermal equilibrium, and its comoving number density must be determined by solving a set of coupled Boltzmann equations (BEQs) (see Sec.~\ref{sec:beq} for details). We assume the mixing between $\phi$ and $h$ is negligibly small, and therefore, $\phi$-mediated scatterings and its decay into SM particles will be highly suppressed resulting not to  consider their effects~\cite{Kaneta:2016vkq, Eijima:2022dec, Seto:2024lik}. As a result, $\phi$—which remains part of the thermal bath—predominantly decays into RHNs and $Z^\prime$ in pairs. To fit the observed DM relic density, it is required that $y_0\, M_{1(3)} = \Omega h^2 \, \frac{1}{s_0}\,\frac{\rho_c}{h^2} \simeq 4.3 \times 10^{-10}\,\text{ GeV}$,
where $y_0 \equiv y_{N_{1(3)}}(z\to\infty)$ is the present DM yield. We use the critical energy density $\rho_c \simeq 1.05 \times 10^{-5}\, h^2$~GeV/cm$^3$, present entropy density $s_0\simeq 2.69 \times 10^3$~cm$^{-3}$~\cite{ParticleDataGroup:2022pth} and DM relic abundance $\Omega h^2 \simeq 0.12$, with $h\simeq H_0/100\,\left({\rm km/s/Mpc}\right)$ being the reduced Hubble rate, where $H_0\simeq 67.4 \pm 0.5 \text{ km/s/Mpc}$ is the current Hubble rate~\cite{Planck:2018vyg}.

%%%%%%%%%%%%%%%
\begin{figure*}[htb!]
\centering        
\includegraphics[scale=.45]{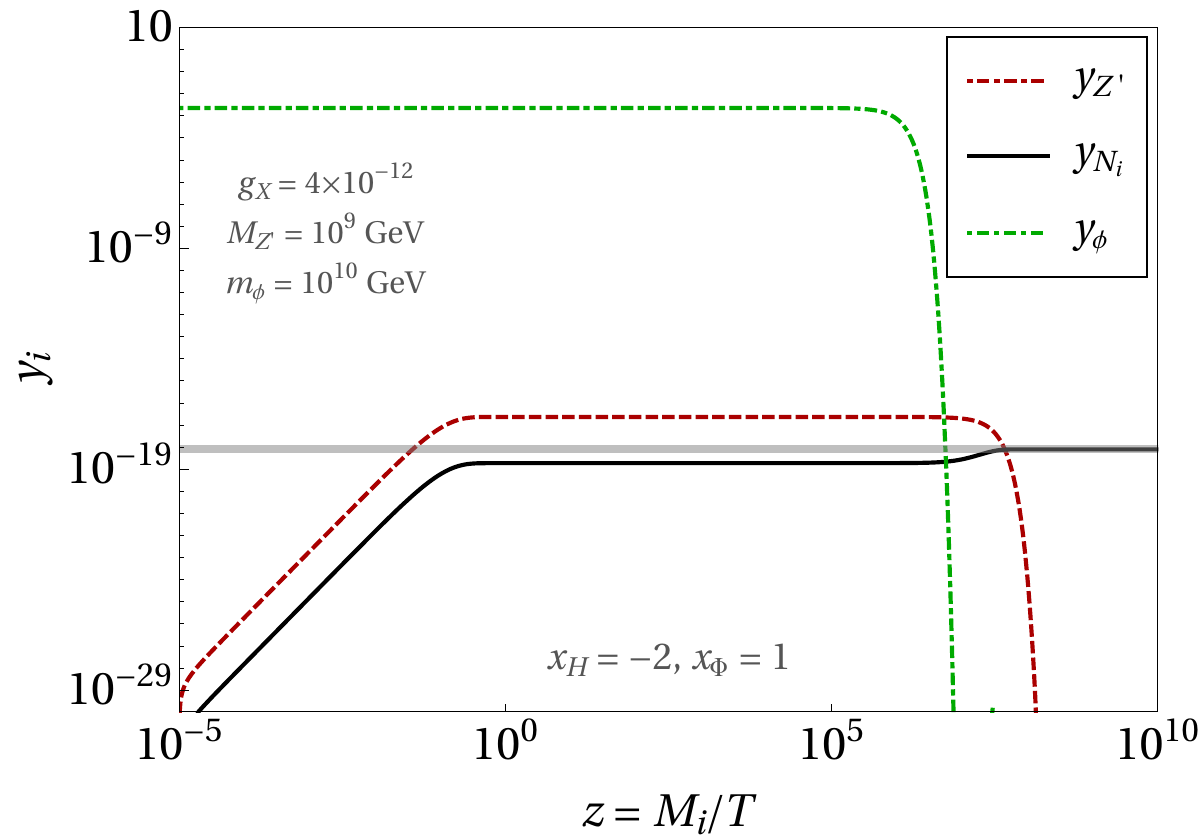}~\includegraphics[scale=.45]{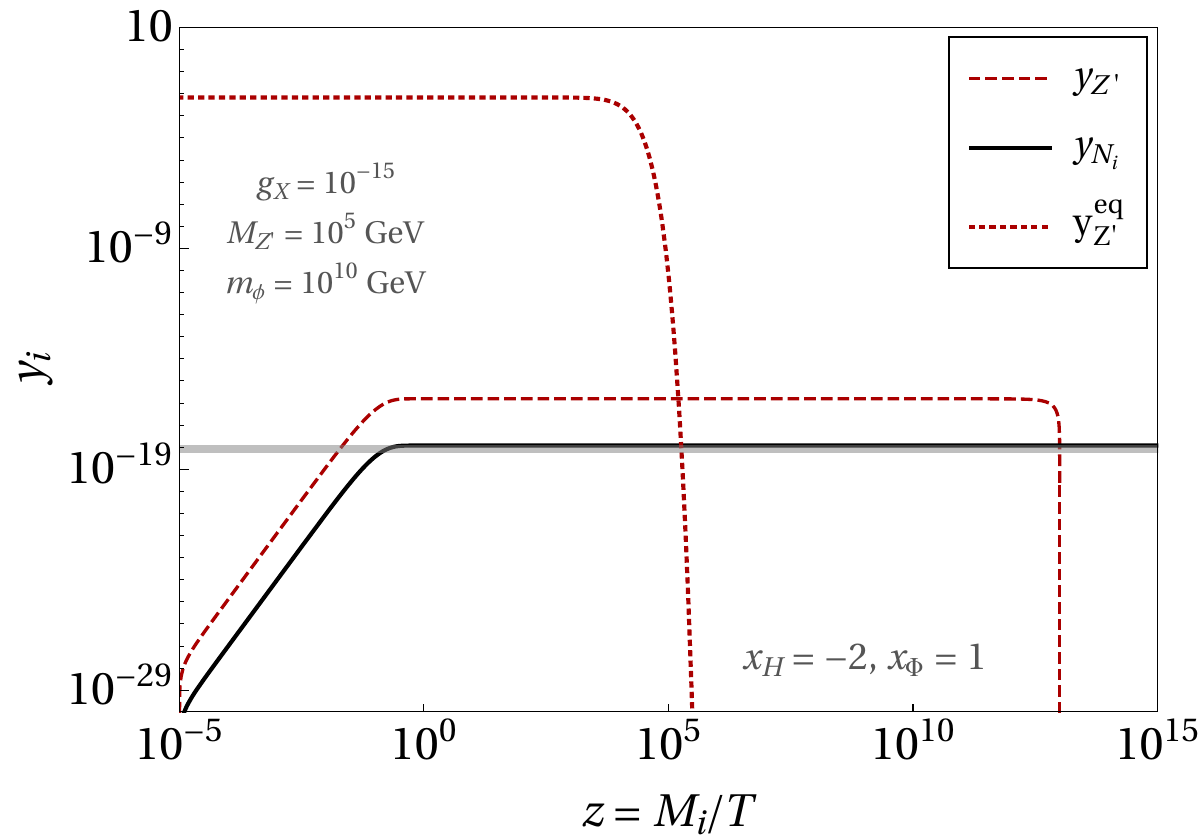}\\[10pt]
\includegraphics[scale=.45]{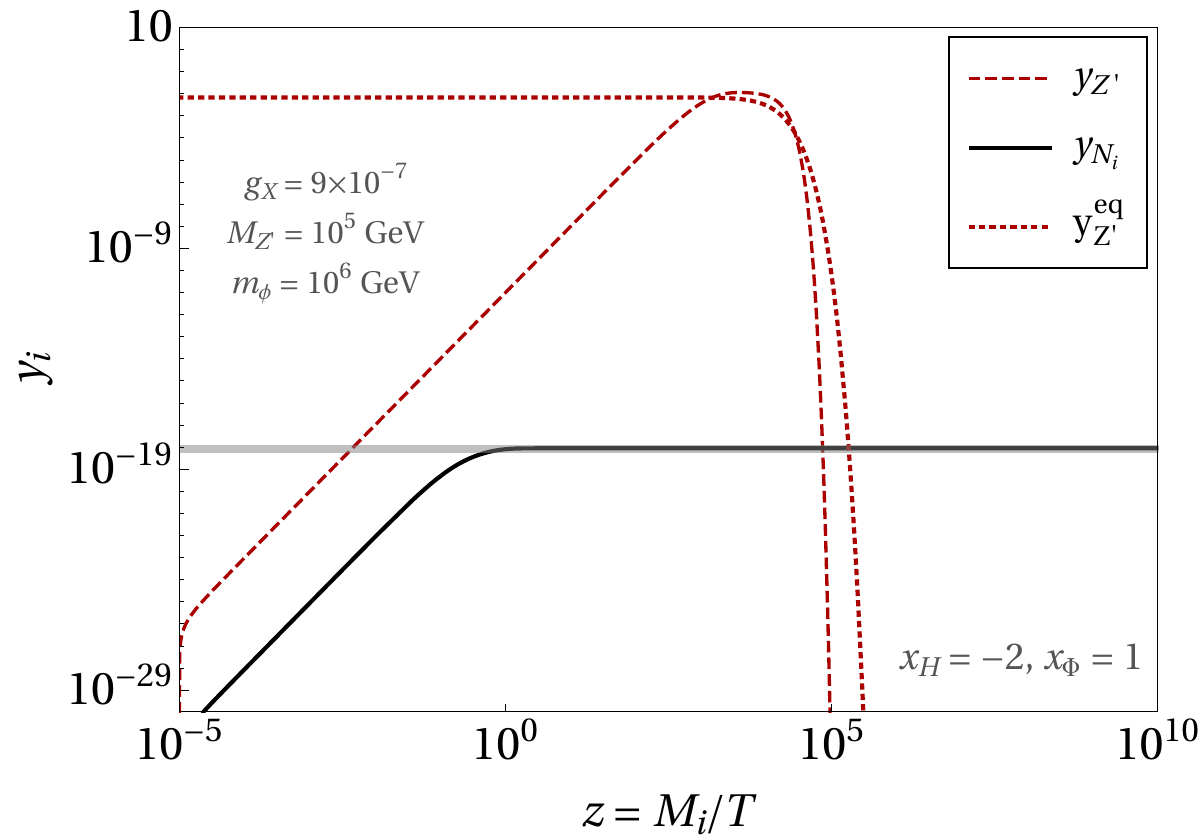}
\caption{Evolution of DM yield (black solid curve), as a function of $z=m_{\nu_\alpha}/T$ with $M_{1(3)}=440$ PeV for NH (IH) case taking $x_H=-2,\,x_\Phi=1$ where DM is dominantly produced from the decays of both $Z'$ and $\phi$ (top left), via $\phi$-decay (top right) and purely from $Z'$-mediated scattering (bottom).}
\label{fig:yld}
\end{figure*}
%%%%%%%%%%%%%%%%%%%
%%%%%%%%%%%%%%%
\begin{figure*}[htb!]
\centering        
\includegraphics[scale=.5]{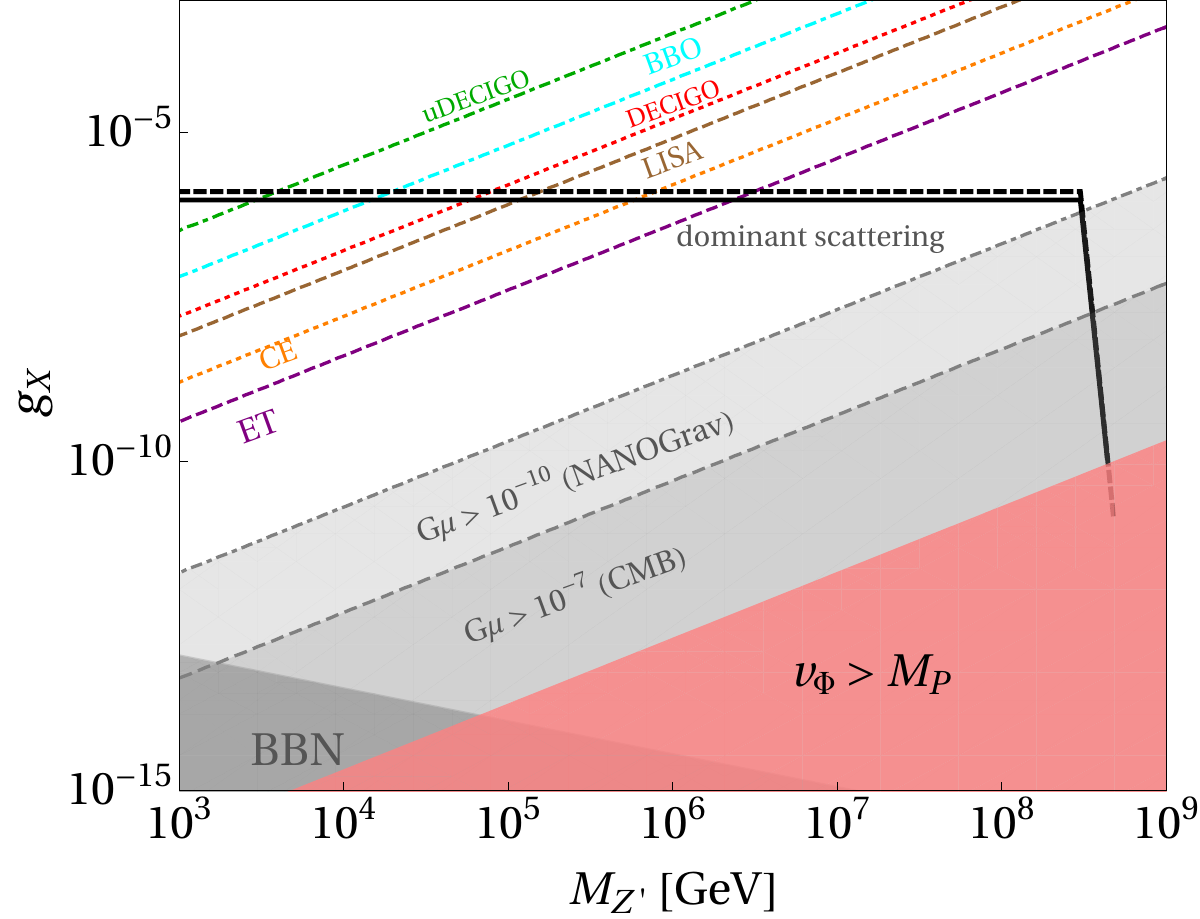}
\caption{Contours of right relic abundance for a 440 PeV DM, shown by the solid and dashed black curves, for $x_H=\{-2,\,0\}$. Along the horizontal lines $Z'$-mediated scattering dominates. The diagonal dashed lines correspond to the sensitivity reaches of several futuristic GW detectors. The shaded regions are disallowed from BBN bound on $Z'$-lifetime, requiring $\tau_{Z'}\gtrsim 1$ sec, Planckian VEV values, CMB and NANOGrav bounds on CS tension. DM production dominantly from $\phi$-decay requires $v_\Phi>M_P$, hence {\it forbidden} in this scenario.}
\label{fig:paramspace}
\end{figure*}
%%%%%%%%%%%%%%%%%%%

The DM yield, as a function of $z=M_{1(3)}/T$ is shown in Fig.~\ref{fig:yld}. In the present scenario, $Z'$ is always produced from the on-shell 2-body decay $\phi\to Z'\,Z'$, therefore, in all cases we consider $m_\phi>2\,\Mzp$. For $M_{1(3)}<\Mzp/2$, as well as $M_{1(3)}<m_\phi/2$, DM is dominantly produced from decays, hence an extremely tiny $g_X$ is required to satisfy the DM abundance. This situation is considered in the top left panel. The DM production shows a non-trivial dynamics here. This is because both $Z'$ and DM yield builds up from $\phi$-decay. Since $\phi$ decays (green dot-dashed curve) earlier compared to $Z'$ (red dashed curve), hence the final DM abundance is set by $Z'$-decay. As a result, we notice a tiny bump at $z\sim 3\times 10^7$, where $Z'$ decay is completed, saturating the DM relic. In the top right panel, DM production dominantly takes place via $\phi$-decay since $M_{1(3)}<m_\phi/2$ but $M_{1(3)}>\Mzp$. Once again, we see, $g_X$ needs to be very feeble to produce the observed DM abundance. It is important to mention that $g_X=10^{-15}$ requires a Planckian $v_\Phi$ in order to satisfy $\Mzp\geq 1$ TeV, and thus forbidden. For DM production entirely via $Z'$-mediated scattering, one requires $M_{1(3)}>\Mzp,\,m_\phi$. Here the DM freezes in at $z\simeq 1$, as shown in the bottom panel. This is the typical IR-feature of freeze-in, where the freeze-in happens at $T={\rm max}\left[\Mzp,\,M_{1(3)}\right]$. Note that, in this case it is possible to have a much larger $g_X$, since scattering channels have a $g_X^4$ dependence, compared to $g_X^2$ dependence for decays.
\\

%%%%%%%%%%%%%%%%%%%%%%%%%%%%%%%%%%%%%%%%%%%%%
\noindent
{\textbf{GW Spectrum from cosmic strings}--} 
%%%%%%%%%%%%%%%%%%%%%%%%%%%%%%%%%%%%%%%%%%%%%%%%
The primary mechanism of energy dissipation from CS is GW emission from oscillating loops, as demonstrated by numerical simulations based on the Nambu-Goto action~\cite{Ringeval:2005kr,Blanco-Pillado:2011egf}. The corresponding energy loss rate is given by~\cite{Vilenkin:1981bx}
\begin{equation}
    P_{\rm GW} = \frac{G}{5} (\dddot{Q})^2\,,
\end{equation}
where $Q$ represents the quadrupole moment of the oscillating loop, and its third time derivative scales as $\dddot{Q} \propto \mu$. Consequently, the energy loss rate follows as
\begin{equation}
    \frac{dE}{dt}=-\Gamma G \mu^2\,,
\end{equation}
where $\Gamma \approx 50$~\cite{Vachaspati:1984gt}. Due to the emission of GWs, the loop undergoes a gradual reduction in length from its initial value $l_i = \alpha t_i$ at the formation time $t_i$, evolving as
\begin{equation}
    l(t)= \alpha t_i - \Gamma G \mu (t - t_i)\,,
\end{equation}
where $\alpha$ denotes the loop size parameter, which simulations suggest is approximately $\alpha \approx 0.1$~\cite{Blanco-Pillado:2013qja,Blanco-Pillado:2017oxo}. The total energy radiated by a loop is distributed among a series of normal mode oscillations, characterized by discrete frequencies
\begin{equation}
    f_k = \frac{2k}{l(t)}\,,
\end{equation}
where $k$ represents the mode number $(k=1,2,3,\dots,\infty)$. The GW spectral density is expressed as
\begin{equation}
    \Omega_{\rm GW}(t_0,f) = \frac{f}{\rho_c} \frac{d\rho_{\rm GW}(t_0,f)}{df} = \sum_k\Omega_{\rm GW}^{(k)}(t_0,f)\,,
    \label{eqn:omgcs1}
\end{equation}
where $f$ and $t_0$ denote the present-day frequency and cosmic time, respectively. Since GW energy density redshifts as $a^{-4}$, we obtain~\cite{Blanco-Pillado:2013qja}
\begin{equation}
    \frac{d\rho_{\rm GW}^{(k)}}{df}=\int_{t_F}^{t_0} \left[\frac{a(t_E)}{a(t_0)}\right]^4 P_{\rm GW}(t_E,f_k) \frac{dF}{df} dt_E\,,
    \label{eqn:omgcs2}
\end{equation}
where $f_k$ denotes the frequency at emission $(f_E)$ at cosmic time $t_E$, with $t_F$ representing the loop formation epoch. The factor $\frac{dF}{df}=f \left[\frac{a(t_0)}{a(t_E)}\right]$ accounts for the redshift of the frequency. The power radiated by the loops is given by
\begin{equation}
    P_{\rm GW}(t_E,f_k) = \frac{2\,k\,G\mu^2\,\Gamma_k}{f\left[\frac{a(t_0)}{a(t_E)}\right]^2} \, n\left(t_E,\frac{2k}{f} \left[\frac{a(t_E)}{a(t_0)}\right]\right)\,,
    \label{eqn:omgcs3}
\end{equation}
where $\Gamma_k$ is defined as
\begin{equation}
    \Gamma_k = \frac{\Gamma k^{-4/3}}{\sum_{m=1}^\infty m^{-4/3}}\,,
\end{equation}
such that $\sum_k\Gamma_k=\Gamma$. The function $n$, number density of loops, depends on the background cosmology characterized by a scale factor $a\propto t^{\beta}$. Using the Velocity-Dependent One-Scale (VOS) model~\cite{Martins:1996jp,Martins:2000cs,Auclair:2019wcv} and numerical simulations~\cite{Blanco-Pillado:2013qja}, the loop number density is given by
\begin{equation}
    n(t_E,l_{k}(t_E))=\frac{A_\beta}{\alpha} \frac{(\alpha+\Gamma G \mu)^{3(1-\beta)}}{\left[l_k(t_E)+\Gamma G \mu t_E\right]^{4-3\beta}t_E^{3\beta}}\,,
    \label{eqn:omgcs4}
\end{equation}
where $A_\beta$ is a constant that depends on the cosmological background. The resulting GW spectrum is influenced by the small-scale structure of loops, which may feature cusps or kinks~\cite{Damour:2001bk,Gouttenoire:2019kij}. Here we assume that cusp-like structures primarily govern the emitted GW spectrum.

Utilizing Eqs.~\eqref{eqn:omgcs1}-\eqref{eqn:omgcs4}, the present-day GW energy density for a given mode $k$ is obtained as
\begin{align}
    \Omega_{\rm GW}^{(k)}(t_0,f)=\frac{2k G\mu^2\Gamma_k}{f \rho_c} \int_{t_{\rm osc}}^{t_0} dt \left[\frac{a(t)}{a(t_0)}\right]^5 n\left(t,l_k\right)\,,
    \label{eqn:omgcsfin}
\end{align}
where the integration extends from $t_{\rm osc}$, the epoch at which loops commence oscillations after being damped by thermal friction~\cite{Vilenkin:1991zk}. This damping phase is subdominant in its effect on the resulting GW spectrum. 

For loops that form and radiate during the radiation-dominated era, the GW spectrum exhibits a characteristic flat plateau, with an amplitude given by
\begin{equation}
    \Omega_{\rm GW}^{(k=1),{\rm plateau}}(f) = \frac{128\pi G\mu}{9\zeta(4/3)} \frac{A_r}{\epsilon_r} \Omega_r \left[(1+\epsilon_r)^{3/2}-1\right]\,,
\end{equation}
where $\epsilon_r = \alpha / \Gamma G\mu$, and $A_r = 0.54$~\cite{Auclair:2019wcv} for the radiation-dominated universe. 
\\

%%%%%%%%%%%%%%%%%%%%%%%%
\noindent
{\textbf{Results and discussions}--} 
%%%%%%%%%%%%%%%%%%%%%
In Fig.~\ref{fig:paramspace} we show the relic density allowed parameter space for a 440 PeV DM. The thick and dashed contours correspond to $x_H=\{-2,\,0\}$, respectively. We show the reach of proposed GW detectors: Big Bang Observer (BBO)~\cite{Crowder:2005nr, Corbin:2005ny}, ultimate DECIGO (uDECIGO)~\cite{Seto:2001qf, Kudoh:2005as}, LISA~\cite{LISA:2017pwj}, the cosmic explorer (CE)~\cite{Reitze:2019iox} and the Einstein Telescope (ET)~\cite{Hild:2010id, Punturo:2010zz, Sathyaprakash:2012jk, Maggiore:2019uih} in probing the parameter space, depending on $v_\Phi$. The light gray shaded region in the bottom left corner corresponds to $\tau_{Z^\prime}=1/\Gamma_{Z^\prime}>1$ sec, that can potentially perturb the predictions of big bang nucleosynthesis (BBN). The darker red shaded region in the bottom right corner demands (super-)Planckian $v_\Phi$. Observations of the Cosmic Microwave Background (CMB) impose an upper limit on the string tension, requiring $G\mu \lesssim 10^{-7}$~\cite{Charnock:2016nzm}. Under this constraint, the condition $\alpha \gg \Gamma G \mu$ holds, leading to $\Omega_{\rm GW}^{(k=1)}(f) \propto v_{\Phi}$. Recent results from NANOGrav~\cite{NANOGrav:2023hvm} impose an even more stringent upper bound, limiting $G\mu \lesssim 10^{-10}$. 

Now, along the horizontal branch of the black curves, the correct DM abundance is produced entirely via $Z'$-mediated scattering. Along these lines, we consider \( m_\phi = 10^6 \) GeV, ensuring \( m_\phi < 2M_{1(3)} \) while maintaining \( m_\phi > 2M_{Z'} \). For \( M_{Z'} \simeq 5 \times 10^7 \) GeV, we set \( m_\phi = 10^9 \) GeV, ensuring that the decay channel \( \phi \to Z' Z' \) is kinematically allowed. Since now \( m_\phi > 2M_{1(3)} \), all decay channels become accessible, thereby requiring an extremely small \( g_X \), as observed in the top panel of Fig.~\ref{fig:yld}. We find $g_X \simeq \{9 \times 10^{-7}, 10^{-6}\}$, corresponding to \( x_H = \{-2, 0\} \) for DM production via \( Z' \)-scattering being independent of neutrino hierarchy as the DM mass is same in both the cases. For dominant DM production from \( \phi \) decay, \( g_X \lesssim\mathcal{O} (10^{-15}) \) is required for \( M_{Z'} \gtrsim 1 \) TeV, which implies $v_\Phi\sim M_P$, thereby excluding this possibility. Since a similar bound is obtained for \( x_H = 0 \), we do not explicitly show this case.  For DM production purely via $Z'$-mediated scattering, we note, in the present framework, KM3NeT provides a slightly stronger bound than IceCube, where latter demands a decaying DM of mass 4 PeV~\cite{Esmaili:2014rma,Higaki:2014dwa}. This can be understood from the fact that since in this case the final DM yield is largely independent of the mediator mass, hence one can write, $\left(g_X^{\rm KM3}/g_X^{\rm ice}\right)\sim\left(4/440\right)^{1/4}\simeq 0.3$, for $x_H=0$.
\\

%%%%%%%%%%%%%%%%%%%%%%%%
\noindent
{\textbf{Conclusions}--} 
%%%%%%%%%%%%%%%%%%%%%
The detection of a flux of UHE neutrinos by the deep-sea neutrino telescope KM3NeT opens a new avenue for investigating high-energy astrophysical sources, both Galactic and extragalactic. Additionally, it offers a powerful tool for constraining physics beyond the SM. This letter presents a minimal particle physics framework in which the origin of these high-energy neutrinos is linked to the decay of a PeV-scale fermionic DM candidate depending on neutrino mass hierarchy. This, in turn, imposes constraints on the mass and coupling of the new neutral gauge boson within the theory. Furthermore, we demonstrate that the viable DM parameter space, satisfying the KM3NeT events, lies within the sensitivity range of multiple gravitational wave detectors, highlighting a valuable complementarity in the quest for searching new physics.
\\

%%%%%%%%%%%%%%%%%%%%%%%%
\noindent
{\textbf{Acknowledgments}-- BB would like to acknowledge fruitful discussions with Suruj Jyoti Das. PS thanks Nayan Das for sharing some data used in this work. The author's thank Latham Boyle for pointing out a crucial typo in the earlier version and for providing a useful reference.}
%%%%%%%%%%%%%%%%%%%%%%%%%%%%%%%%%%%%%%%%%%%%
%\clearpage
%\onecolumngrid
%%%%%%%%%%
\begin{widetext}
\appendix
\section{Neutrino mass}
\label{sec:nu-mass}
%%%%%%%%%%%%%%%%%%%%%%%%%%%%
In order to generate the neutrino mass from the seesaw mechanism, the neutrino mass matrix can be written as \bea
{\cal M}_{\nu}=\begin{pmatrix}
0&&M_{D}\\
M_{D}^{T}&& M_N
\end{pmatrix}.
\label{typeInu}
\eea
Assuming the hierarchy of $|M_{D_{ij}}/M_{N_\beta}| \ll 1$, we diagonalize the neutrino mass matrix to obtain the non-zero neutrino mass eigenvalue  of light Majorana neutrino as
\bea
m_{\nu} \simeq - M_{D} M_{N}^{-1} M_{D}^{T}.
\label{seesawI}
\eea
which is the well known seesaw formula. Due to the light-heavy neutrino mixing from the seesaw mechanism, a flavor eigenstate of light neutrino ($\nu_\alpha$) can be written as a linear combination of the mass eigenstates of light ($\nu_i$) and heavy ($N_i$) neutrino
\bea 
  \nu_\alpha \ \simeq \ \mathcal{U}_{\alpha i} \nu_i  + V_{\alpha i} N_i \, ,  
\eea 
where $\mathcal{U}$, the PMNS matrix, is taken at the leading order after ignoring the non-unitarity effects for simplicity. Now we diagonalize the light neutrino mass matrix as 
\bea
\mathcal{U}^T m_\nu \mathcal{U} ={\rm diag}(m_1, m_2, m_3).
\eea
where PMNS matrix can be given by 
\bea
\mathcal{U}= \begin{pmatrix} C_{12} C_{13}&S_{12}C_{13}&S_{13}e^{i\delta}\\-S_{12}C_{23}-C_{12}S_{23}S_{13}e^{i\delta}&C_{12}C_{23}-S_{12}S_{23}S_{13}e^{i\delta}&S_{23} C_{13}\\ S_{12}C_{23}-C_{12}C_{23}S_{13}e^{i\delta}&-C_{12}S_{23}-S_{12}C_{23}S_{13}e^{i\delta}&C_{23}C_{13} \end{pmatrix} \begin{pmatrix} 1&0&0\\0&e^{i\rho_1}&0\\0&0&e^{i \rho_2}\end{pmatrix}
\label{pmns}
\eea
with $C_{ij}=\cos\theta_{ij}$, $S_{ij}=\sin\theta_{ij}$, $\delta (\rho_{1,2})$ is the Dirac(Majorana) $CP$ phase. In this analysis we adopt neutrino oscillation data as $\sin^{2}2{\theta_{13}}=0.092$,
$\sin^2 2\theta_{12}=0.87$, $\sin^2 2\theta_{23}=1.0$, $\Delta m_{12}^2 = m_2^2-m_1^2 = 7.6 \times 10^{-5}$ eV$^2$, and $|\Delta m_{23}^2|= |m_3^2-m_2^2|=2.4 \times 10^{-3}$ eV$^2$ from \cite{ParticleDataGroup:2024cfk} for the NH and IH cases. Now we write the mixing between the light and heavy neutrino mass eigenstates in the seesaw scenario following general parametrization as  
\bea
V^{\rm{NH/IH}} \ = \mathcal{U}_{}^{\ast} \sqrt{D^{\rm{NH/ IH}}} \, \mathcal{O} \, \sqrt{M_N^{-1}},
\label{gp2}
\eea
where $\mathcal{O}$ is a general orthogonal matrix: 
\bea
\mathcal{O}\ = \ 
\begin{pmatrix}
1&0&0\\
0&\cos x& \sin x\\
0&-\sin x& \cos x
\end{pmatrix}
\begin{pmatrix}
\cos y&0&\sin y\\
0&1& 0\\
-\sin y& 0&\cos y
\end{pmatrix}
\begin{pmatrix}
\cos z&\sin z&0\\
-\sin z&\cos z&0\\
0&0&1
\end{pmatrix}
\label{Omatrix}
\eea
with the angles, $x, y, z$ being complex numbers, and $D_{\rm NH/IH}$ is the light neutrino mass eigenvalue matrix
\bea 
  D^{\rm{NH}} \ = \ {\rm diag}
  \left(m_{\rm lightest}, m_2^{\rm{NH}}, m_3^{\rm{NH}} \right),  
 \,\,\,\,\,\,  D^{\rm{IH}} \ = \ {\rm diag}
\left( m_1^{\rm{IH}}, m_2^{\rm{IH}}, m_{\rm lightest} \right) 
\label{DNH}
\eea 
with $m_2^{\rm{NH}}=\sqrt{ \Delta m_{12}^2+m_{\rm lightest}^2}$, $m_3^{\rm{NH}}=\sqrt{\Delta m_{23}^2 + (m_2^{\rm{NH}})^2}$, $m_2^{\rm{IH}}=\sqrt{ \Delta m_{23}^2 + m_{\rm lightest}^2}$ 
and $m_1^{\rm{IH}}=\sqrt{(m_2^{\rm{IH}})^2- \Delta m_{12}^2}$. In both cases, the RHN mass matrix is defined as  $M_{N} \ = \ {\rm diag}\left(M_{N_{1}},  M_{N_{2}}, M_{N_{3}} \right)$
Hence, the mixing matrix $V$ in Eq.~\eqref{gp1} becomes a function of $\rho_{1,2}$, $m_{\rm lightest}$, $M_{N_\beta}$ ($\beta=1,2,3$), the three complex angles and neutrino oscillation data. For a simple parametrization one can consider that $x,y,z=0$ making $\mathcal{O}$ an identity matrix. Now the modified charged-current (CC) interaction in the lepton sector can be written as 
\bea 
\mathcal{L}_{\rm CC} \ = \ 
 -\frac{g}{\sqrt{2}} W_{\mu}
  \bar{\ell} \gamma^{\mu} P_L 
   \left[\mathcal{U}_{\alpha i} \nu_i+ V_{\alpha i} N_i\right] + {\rm H. c.}\,, 
\label{CC}
\eea
where $g$ is the $SU(2)_L$ gauge coupling and the modified neutral-current (NC) interaction in the lepton sector will be
\bea 
\mathcal{L}_{\rm NC} = -\frac{g}{2 \cos\theta_w}  Z_{\mu} \left[ (\mathcal{U}^\dag \mathcal{U})_{ij} \bar{\nu}_i \gamma^{\mu} P_L \nu_j + (\mathcal{U}^\dag V)_{ki} \bar{\nu}_k \gamma^\mu P_L N_i + (V^\dagger V)_{mi} \bar{N}_m\gamma^\mu P_L N_i\right] + {\rm H. c.}\,, 
\label{NC}
\eea
where $\theta_w$ is the weak mixing angle. 
%%%%%%%%%%%%%%%%%%
\section{Interactions and decay widths}
\label{sec:decay}
%%%%%%%%%%%%%%%%%%
Under the general $U(1)_X$ scenario left- and right-handed fermions interact differently with the $Z^\prime$ and the interaction Lagrangian manifest chiral scenario involving their general $U(1)_X$ charges. We write the interactions Lagrangian as
\bea
\mathcal{L} = -g_X (\overline{f}\gamma^\mu q_{f_{L}^{}}^{} P_L^{} f+ \overline{f}\gamma^\mu q_{f_{R}^{}}^{}  P_R^{} f) Z_\mu^\prime~,
\label{Lag1}
\eea
where $P_{L(R)}^{}= (1 \mp \gamma_5)/2$ is the left- (right-)handed projections and 
$q_{f_{L(R)}^{}}^{}$ is the corresponding general $U(1)_X$ charge of the left- (right-)handed fermion $(f_{L(R)}^{})$. The partial decay widths of $Z^\prime$ into a pair of charged fermions can be written as
\begin{align}
\label{eq:width-ll}
    \Gamma(Z' \to \bar{f} f)
    &= N_C^{} \frac{M_{Z^\prime}^{} g_{X}^2}{24 \pi} \left[ \left( q_{f_L^{}}^2 + q_{f_R^{}}^2 \right) \left( 1 - \frac{m_f^2}{M_{Z^\prime}^2} \right) + 6 q_{f_L^{}}^{} q_{f_R^{}}^{} \frac{m_f^2}{M_{Z^\prime}^2} \right] \left( 1 - 4 \frac{m_f^2}{M_{Z^\prime}^2} \right)^{\frac{1}{2}}~,
    \end{align}    
where $m_f$ is the mass of the SM fermions and $N_C^{}=1(3)$ is the color factor for the SM leptons (quarks). For heavy $Z^\prime$ ion the TeV scale and above, the effect of the SM masses can be neglected. The partial decay width of $Z^\prime$ into a pair of light neutrinos $(\nu_L)$ can be written as
\begin{align}   
\label{eq:width-nunu}
    \Gamma(Z' \to \nu \nu)
    = \frac{M_{Z^\prime}^{} g_{X}^2}{24 \pi} q_{\ell_L^{}}^2~,
\end{align} 
neglecting the tiny light neutrino mass and $q_{\ell_L}$ is the $U(1)_X$ charge of the SM lepton doublets. The $Z^\prime$ gauge boson can decay into a pair of heavy Majorana neutrinos if $M_{Z^\prime} > M_{N_\beta}/2$. The kinematically allowed partial decay width of $Z^\prime$ into one generation of heavy neutrino pair can be given by
\begin{align}
\label{eq:width-NN}
    \Gamma(Z^\prime \to N_\beta N_\beta)
    = \frac{M_{Z^\prime}^{} g_{X}^2}{24 \pi} q_{N_R^{}}^2 \left( 1 - 4 \frac{M_{N_\beta}^2}{M_{Z^\prime}^2} \right)^{\frac{3}{2}}~,
\end{align}
where $q_{N_R^{}}^{}$ is the general $U(1)_X$ charge of the heavy neutrinos. If we consider the scenario where $M_{N_\beta} > M_{Z^\prime}/2$, then $Z^\prime \to N_\beta N_\beta$ mode will be absent. In addition we write the dominant decay rates of the BSM scalar $\phi$ into a pair of $Z^\prime$ from the kinetic term and into a pair of RHNs from the Yukawa interaction as
\bea
\Gamma_{\phi\to Z^\prime\,Z^\prime}=\frac{g_X^2}{8 \pi\,r^4}\,\frac{\Mzp^2}{m_\phi}\,\sqrt{1-4\,r^2} \left(12\,r^4-4\,r^2+1\right)
\eea
\bea
\Gamma_{\phi\to N_\beta\,N_\beta}=\frac{g_X^2}{2\pi}\,\left(\frac{M_{N_\beta}}{\Mzp}\right)^2\,m_\phi\,\left(1-\frac{4\,M_{N_\beta}^2}{m_\phi^2}\right)^{3/2}
\eea
where $r=M_{Z^\prime}/m_\phi$. 
%%%%%%%%%%%%%%%%%%%%
%%%%%%%%%%%%%%%%%%%%%%%%%%%%
\section{DM production Cross-section}
\label{sec:cs}
%%%%%%%%%%%%%%%%%%%%%%%%%%%%
For charged lepton initial states,
\begin{align}
& \sigma(s)_{\ell^+\ell^-\to\text{DM}\text{DM}}\simeq \frac{g_X^4}{96\pi\,\left[(s-\Mzp^2)^2+\Gamma_{Z'}^2\,\Mzp^2\right]}\,\sqrt{1-\frac{4\,\mdm^2}{s}}\,
x_\Phi^2\,\frac{\left(s-4\,\mdm^2\right)\,\left(5\,x_H^2+12\,x_H\,x_\Phi+8\,x_\Phi^2\right)}{2}\,.
\end{align}
For light neutrino initial states
\begin{align}
& \sigma(s)_{\nu\nu\to\text{DM}\text{DM}}\simeq \frac{g_X^4}{48\pi\,\left[(s-\Mzp^2)^2+\Gamma_{Z'}^2\,\Mzp^2\right]}\,\sqrt{1-\frac{4\,\mdm^2}{s}}\,x_\Phi^2\,\frac{\left(s-4\,\mdm^2\right)\,\left(x_H+2\,x_\Phi\right)^2}{4}
\end{align}
For up-quark initial states,
\begin{align}
& \sigma(s)_{uu\to\text{DM}\text{DM}}\simeq x_\Phi^2\,\frac{\left(s-4\,\mdm^2\right)\,\left(17\,x_H^2+20\,x_H\,x_\Phi+8\,x_\Phi^2\right)}{2}\,.
\end{align}
For down-quark initial states,
\begin{align}
& \sigma(s)_{dd\to\text{DM}\text{DM}}\simeq x_\Phi^2\,\frac{\left(s-4\,\mdm^2\right)\,\left(5\,x_H^2-4\,x_H\,x_\Phi+8\,x_\Phi^2\right)}{2}\,.
\end{align}
%%%%%%%%%%%%%%%
\section{Boltzmann equations}
\label{sec:beq}
%%%%%%%%%%%%%%%
The coupled BEQs for freeze-in production of the DM read,
\bea
\frac{dy_\phi}{dz}&=&-\frac{z}{\mathcal{H}}\,\langle \Gamma_{\phi}\rangle\,y_{\rm eq}^\phi + \frac{s}{\mathcal{H}}\, \frac{1}{z^2}\,\langle\sigma v\rangle_{\text{SM}\, \text{SM}\to \phi\,\phi}\,\left(y_\text{eq}^\phi\right)^2, \nonumber \\
\frac{dy_{Z^\prime}}{dz}&=& - \frac{z}{\mathcal{H}}\langle \Gamma_{Z^\prime}\rangle\, y_{Z^\prime} +\frac{z}{\mathcal{H}}\langle \Gamma_{\phi \to Z^\prime Z^\prime }\rangle\,\left(y_\text{eq}^\phi-y_{Z^\prime}\right),
\nonumber\\
\frac{dy_{N_1}}{dz}&=&\frac{z}{\mathcal{H}}\,\langle \Gamma_{Z^\prime \to N_1N_1}\,\rangle y_{Z^\prime}+ \frac{z}{\mathcal{H}}\,\langle \Gamma_{\phi\to N_1N_1}\rangle\,y_{\rm eq}^\phi + \frac{s}{\mathcal{H}}\, \frac{1}{z^2}\,\langle\sigma v\rangle_{\text{SM}\, \text{SM}\to N_1 N_1}\,y_\text{eq}^2\,, 
\label{eq:cBEQ}
\eea
where $y_i\equiv n_i/s$ is the yield of a certain species $i$, with
\begin{align}
& y_j^\text{eq} = \frac{45}{4\,\pi^4}\,\frac{g_j}{\gss}\,z^2\,K_2[z]\,,    
\end{align}
is the equilibrium yield, with $g_j$ being the degrees of freedom for the corresponding $j$ particle and $z=M_1/T$ is a dimensionless variable. Here, $\mathcal{H}=(\pi/3)\,\sqrt{\gs/10}\,\left(T^2/M_P\right)$ is the Hubble parameter for a standard radiation dominated (RD) Universe and $s=\left(2\pi^2/45\right)\,\gss(T)\,T^3$ is the entropy density. The number of relativistic degrees of freedom in the bath corresponding to energy density and entropy density are tracked by $\gs(T)$ and $\gss(T)$, respectively. At temperatures well above the QCD phase transition we have $\gs(T)\simeq\gss(T)\approx 106$.The thermally averaged decay rate is given by
\begin{align}
& \langle\Gamma_{\phi\to jj}\rangle  = \frac{K_1(z)}{K_2(z)}\times\Gamma_{\phi\to jj}\,,  
\end{align}
with $z=M_1/T$, and $j$ represents the final state particle.
%%%%%%%%%%%%%%%%%%%%%%%%%%%
\end{widetext}
\vspace{-0.398in}
\bibliographystyle{utphys}
\bibliography{bibliography}
\end{document}